# Simulation of optical microfiber loop resonators for biochemical sensing

**Lei Shi and Xianfeng Chen**

*Institute of Optics and Photonics, Department of Physics, Shanghai Jiao Tong University, 800 Dongchuan Road, Shanghai 200240, China*

**Abstract:** Based on the basic theory of the microfiber loop resonator, we exploit the application of microfiber loop resonators in biochemical sensing. We set up a reliable theoretical model and optimize the structural parameters of microfiber loop resonators including the radius of the microrfiber, the radius of the loop and the length of the coupling region for higher sensitivity, wider dynamic measurement range, and lower detection limit. To show the convincible and realizable sensing ability we perform the simulation of sensing an extreme small variation of ambient refractive index by employing a set of experimental datas as the parameters in the expression of intensity transmission coefficient, and the detection limit reaches to a varation of ambient refractive index of $10^{-5}$ refractive index unit(RIU). This has superiority over the exsiting evanescent field-based subwavelength-diameter optical fiber refractive index sensor.

---

### References and links

---

## 1. Introduction

Recently optical waveguides with subwavelength diameter attract much attention, especially optical microfiber-based photonic devices[1-6]. Fabrication methods of ultra-high-quality silica optical microfibers with length up to ~10cm and diameter down to ~30nm have been developed[2-3,6], so silica optical microfibers are promising to develop micro- and nanophotonic devices. Demonstrated and potential applications of microfibers include nonlinear optics[3,7-8], optical sensing[9-11], and the microfiber loop resonator(MLR)[12-17].Compared to integrated optical waveguide ring resonators[18], MLRs are more convenient for fabrication, manipulation and application. An MLR can be directly obtained using a microfiber manipulated into a loop, where effective coupling occurs between the contact sections of the microfiber. The MLR with loaded *Q*-factor of 120000 has been produced[16], and the performance is close to that of integrated waveguide microring resonators of which the largest *Q*-factor has been demonstrated recently[19].

MLRs are also potential useful for optical filtering, active devices and optical sensing as the integrated waveguide microring resonators[18]. However, the research on MLRs for biochemical sensing hasn't been reported. In this paper, we simulate the response of MLRs to the variation of ambient refractive index and optimize the structural parameters of MLRs to provide a theoretical guide for variant requirements in practical sensing applications. After that we perform the simulation based on a set of experimental datas in reference[16] to show the achievable sensing perfomance of MLRs.

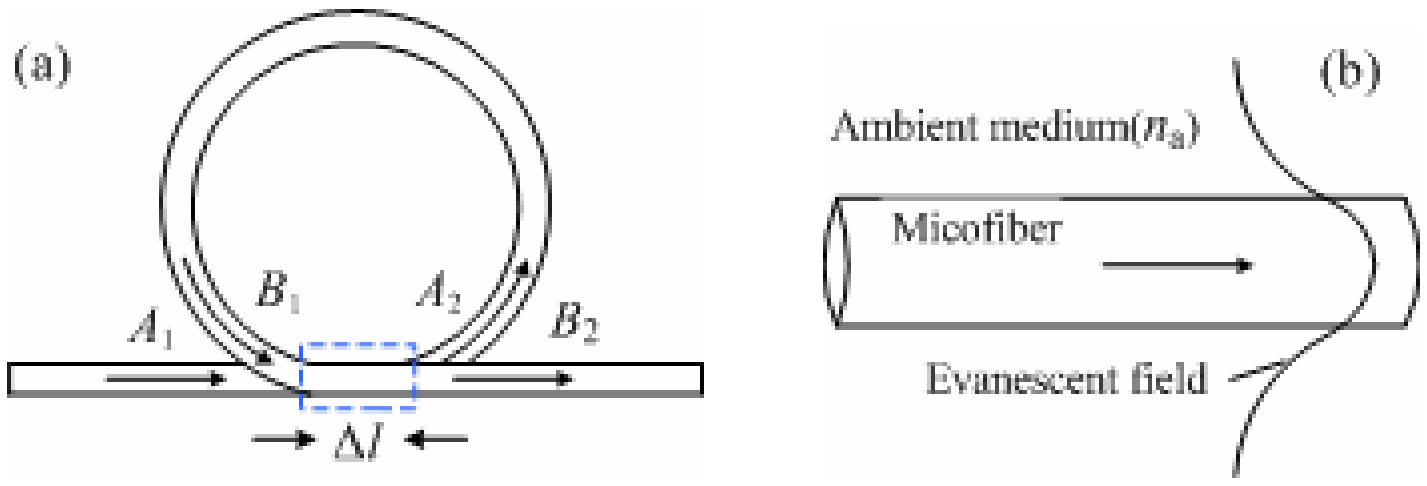

Fig. 1(a) Geometry of of an MLR and (b) Optical signal transmits in a microfiber with the ambient refractive index $n_a$.

## 2. Theory

### A. Fundamental s of an MLR

A typical MLR is shown in Fig. 1(a). $A_1$, $A_2$, $B_1$ and $B_2$ are the complex mode amplitudes. Coupling of optical singal occurs in the region selected by the blue dashed line frame, where the adjacent microfibers contact and parallel with each other. The loop can be treated as a cirlce, and then the length of the loop $L$ is equal to $2\pi R$, where $R$ is the radius of the loop.

For simplicity the polarization effect is ignored[16]. Based on the fact that fabrication of ultra-low-loss silica microfibers with length up to ~10cm and uniform diameter has been demonstrated[2-3,6], we can assume that the field distribution in the loop is independent on the specific axial position for the theoretical simulation. That is to say, both propagation constant $\beta$ and coupling coeffcient $\kappa$ are constant funtions of the axial position.

The intensity transmission coeffcient $T$, representing the ratio of the transimission intensity to the input intensity, can be obtained according to general principle of ring resonators [16,22-23]and has the expression[16]

$$T = \left|\frac{B_2}{A_1}\right|^2 = \left|\frac{\exp(-\frac{\alpha L}{2})\exp(i\varphi) - \sin(\mathrm{K})}{1-\sin(\mathrm{K})\exp(-\frac{\alpha L}{2})\exp(i\varphi)}\right|^2 = \frac{\exp(-\alpha L) + (\sin(\mathrm{K}))^2 - 2\exp(-\alpha L/2)\sin(\mathrm{K})\cos(\varphi)}{1+\exp(-\alpha L)(\sin(\mathrm{K}))^2 - 2\exp(-\alpha L/2)\sin(\mathrm{K})\cos(\varphi)} \tag{1}$$

where $\alpha$ is the intensity attenuation coeffcient (*viz*. power attenuation coeffcient for this model), $L$ is the length of the loop, $\varphi$ is the roundtrip phase shift, $K$ is the coupling parameter. For the field distribution in the microfiber is independent on the axial position we can get $\varphi=\beta L$ and $K=\kappa\Delta l$, where $\Delta l$ is the length of the coupling region.

From the expression of $T$, we know the resonances in transmission spectrum generate when $\cos(\varphi)$ is equal to zero and the coupling parameter $K$ is close to $K_n=(2n+1)\pi/2$, where $n$ is an integer [16,22-23].

The free spectrum range(FSR) of an MLR, which is defined as the spacing between adjacent resonant wavelengths, and the $Q$-factor of an MLR, which is defined as of resonant wavelength $\lambda$ to the full width at half maximum of the corresponding resonance spectrum, can be respectively expressed as when $\alpha L \ll 1$ [16]

$$Q = \frac{\lambda L}{(K-K_n)^2 + \alpha L}\cdot\frac{\partial\beta}{\partial\lambda},\quad FSR = \frac{2\pi}{L}\cdot\frac{1}{\frac{\partial\beta}{\partial\lambda}} \tag{2}$$

**B. Sensing parameters**

The sensing mechanism of MLR-based biochemical sensors is based on the shift of resonant wavelength $\lambda_r$ due to a variation of ambient refractive index caused by varied ambient material concentration, and it is illustrated in Fig.1(b). The microfiber is so thin that fractional optical power transimits in the ambient cladding [9]. As the refractive index of ambient material $n_a$ changes the effective index of the guided modes in the microfiber $n_{eff}$ is modified, and then the resonant wavelength shifts.

Following the sensing mechanism the sensitivity can be defined as[20]

$$S = \frac{\partial\lambda_r}{\partial n_a} = \frac{\partial\lambda_r}{\partial n_{eff}}\cdot\frac{\partial n_{eff}}{\partial n_a} \tag{3}$$

Another important parameter for sensing applications is the detection limit $\delta n_a$. According to the definition of sensitivity the detection limit has the feature

$$\delta n_a \sim \delta\lambda_r \cdot \frac{\partial n_a}{\partial \lambda_r} \to \delta n_a \propto \frac{1}{S} \tag{4}$$

where $\delta\lambda_r$ is the spectral resolution of the measurement system. Theoretically the higher $S$ is, the smaller the detection limit is. However, in fact it is very difficult to exactly obtain the shift of the resonant wavelength if the spectrum width is large, for the experimental line shape usually doesn't agree with the theoretical line shape so well. So only if the $Q$-factor is high enough the low detection limit can be achievable, and we define the detection limit factor $P = Q \cdot S$[20] to characterize the detection limit. The larger $P$ is, the lower the detection limit is.

**3. Optimization of the structural parameters of the MLR for ambient refractive index sening**

From (2), (3) and (4) we know that the sensing parameters sensitivity and detection limit are determined by several key structural parameters of the MLR: the radius of the microfiber $a$, the radius of the loop $R$ and the length of the coupling region $\Delta l$, so we perform our simulations to optimize these parameters for high-performance sensing.

The operation wavelength is set in ~1.55μm, and we choose the radius of microfibers ranging from 0.3μm to 0.55μm to satisfy single mode operation and low loss for our simulations.

For the instrinsic loss of microfibers is ultra low we only consider the attenuation due to bend loss for $\alpha$[21]. $\kappa$ can be obatained from the coupled-mode theory[16,21].

At first we pay attention to the sensitivity. From (3) the sensitivity is related to the raidus of the microfiber and the loop. The results shown in Fig.2(a) indicate that the sensitivity increases as the raidus of the microfiber decreases, and hardly changes with the raidus of the loop. Microfibers with smller raidus have higher fractional power transimitting in the ambient cladding, and then are more sensitive to the variation of ambient refractive index. As to the radius of the loop, $\partial n_{eff}/\partial n_a$ in (3) is independent on $R$, and $\partial\lambda_r/\partial n_{eff}$ in (3) is hardly dependent on $R$, which can be understood from the first resonacne condition. Moreover, from(4) the theoretical detection limit is a variation of ~$10^{-6}$ RIU based on an optical spectrum analyser with the resolution of 1pm as the measurement apparatus.

For sensing applications, wide dynamic measurement range is also important, and it is related to the FSR based on this sensing mechanism. From the definition FSR is independent on the coupling parameter, and its evolution trend is shown in Fig. 2(b). Small $a$ and $R$ are necessary to obtain larger FSR, and $R$ is the dominant factor ,which can be understood from the definition of FSR.

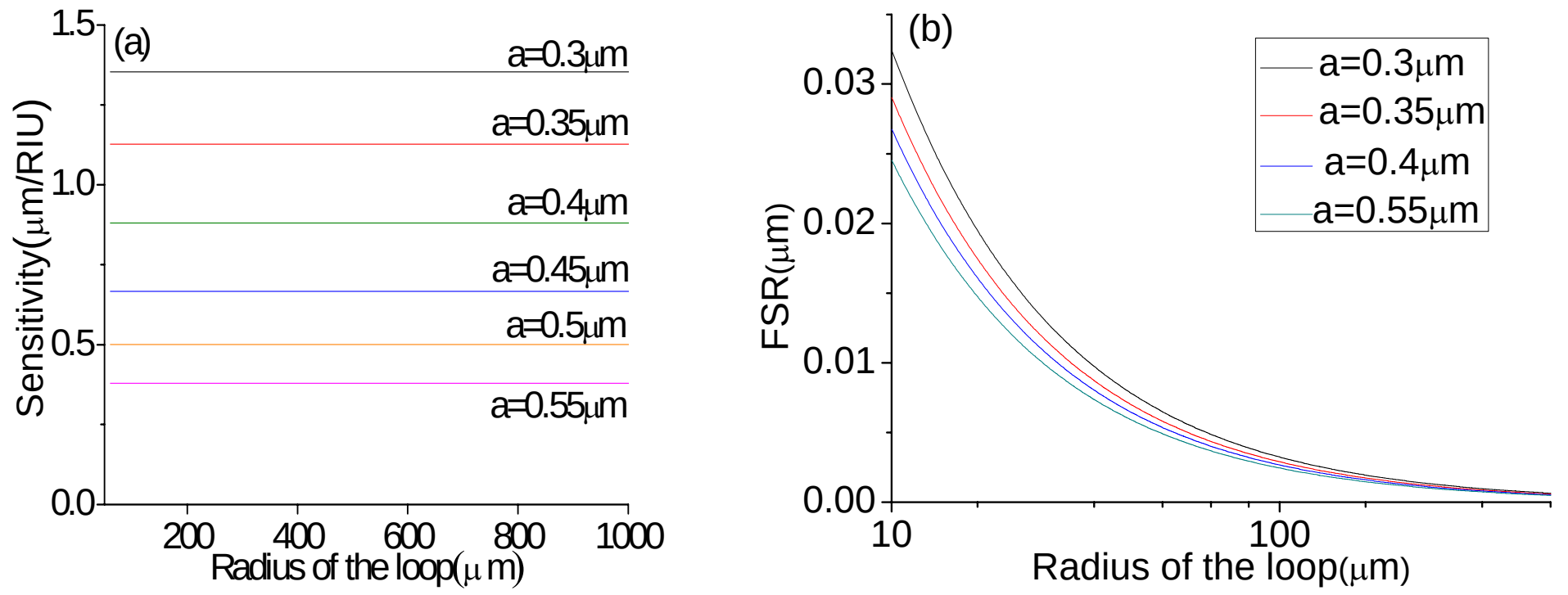

Fig. 2 (a) Sensitivity versus the radius of the loop *R* and the radius of the microfiber *a*, the unit of sensitivity is μm per refractive index unit (RIU) (b) FSR versus the radius of the loop *R* and the radius of the microfiber *a*.

*Q*-factor is the most important parameter for a resonator for it characterizes the perfomance of a resonator. Though the asymptotic formula of power attenuation coeffcient $\alpha$ isn't so accurate when the radius of the loop is very small (e.g. $a$~10μm)[4], it is still reliable to simulate the evolution trend of the *Q*-factor and the detection limit factor *P* as the structural parameters of the MLR change.

From (3) we know *Q*-factor depends on *a*, *R* and $\Delta l$. The resluts shown in Fig.3(a) indicate that higher *Q*-factor corresponds to large *R* and large *a*. In fact the *Q*-factor linearly depends on *R* if we use conventional linear scale instead of logarithm scale, and this is similar to integrated waveguide microring resonators [24]. High *Q*-factor also can be obtained based on large *a* for corresponding lower loss when *R* is large. The coupling length $\Delta l$ is another important parameter that affects the *Q*-factor. The coupling parameter *K* is periodic function of $\Delta l$, and $(K-K_n)^2$ is the same. So *Q*-factor is modified periodically as the coupling length variates. We come to a conclusion that the coupling length $\Delta l$ which makes *K* close to $K_n$ is necessary to obtain higher *Q*-factor.

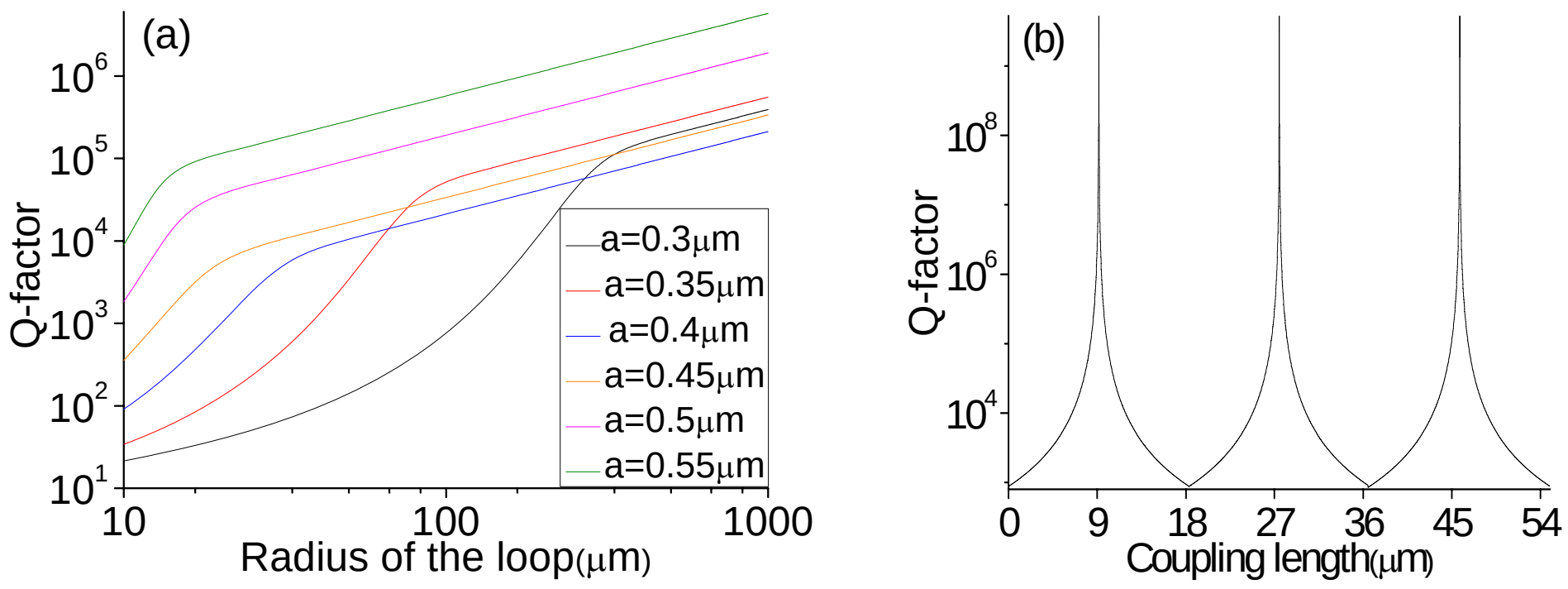

Fig. 3 (a) $Q$-factor versus the radius of the loop $R$ and the radius of the microfiber $a$, fixed parameter $\Delta l$=10μm (b) $Q$-factor versus the coupling length $\Delta l$, fixed parameters $a$=0.5μm, $R$=55μm.

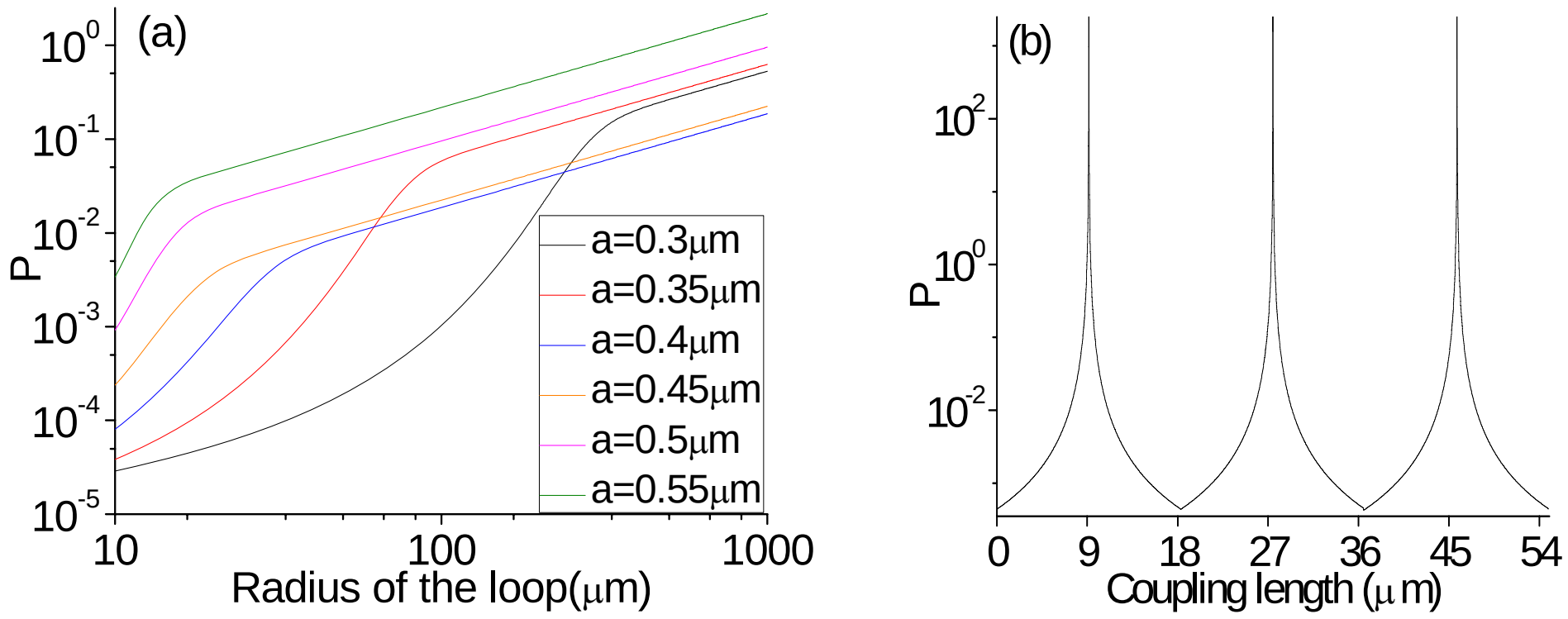

Fig. 4 (a) The detection limit factor P versus the radius of the loop R, fixed parameter $\Delta l$=10μm. (b) The detection limit $P$ versus the coupling length $\Delta l$, fixed parameters $a$=0.5μm, $R$=55μm.

At last the detection limit factor is caculated. It is shown in Fig.4 that the evolution trend of detection limit factor is similar to that of $Q$-factor. Lower detection limit can be obtained by choosing large $a$, large $R$ and appropriate $\Delta l$. According to the preceding discussion the detection limit is in the inverse ratio of both $Q$-factor and the sensitivity. However, these two parameters have inverse dependences on the radius of the microfiber, so we should to find a balance between them. The results indicate that the detection limit factor has similar trend to that of the $Q$-factor. That is to say, the $Q$-factor is the dominant parameter for affecting the detection limit.

Based on the foregoing simulations we find the structural parameters of MLRs which correspond to better sensing performance following our variant practical requirements.

## 4. Simulation of sensing an extreme small variation of ambient refractive index

To show the convincible and realizable sensing ability we perform the simulation employing a set of experimental datas in reference[16] as the parameters in equation (1). The radius of the microfiber is 0.45μm and the length of the loop is 2mm. Corresponding roundtrip attenuation $\alpha L$ is 0.14 and amplitude coupling coefficient sin($K$) is 0.981 with operation wavelength ~1.55μm [16]. The resolution of an optical spectrum analyser is 1pm, so we can obtain that the detection limit of this MLR-based sensor is an ambient refractive index variation of $10^{-5}$ RIU from the simulation results shown in Fig.5. It is indicated that the MLR-based sensor has superiority over the exsiting evanescent field-based subwavelength-diameter optical fiber refractive index sensor of which the detection limit is a variation of ~$10^{4}$RIU[11]. As the improvement of fabrication technology of MLRs, we believe the theoretical detection limit of ~$10^{-6}$RIU is achievable.

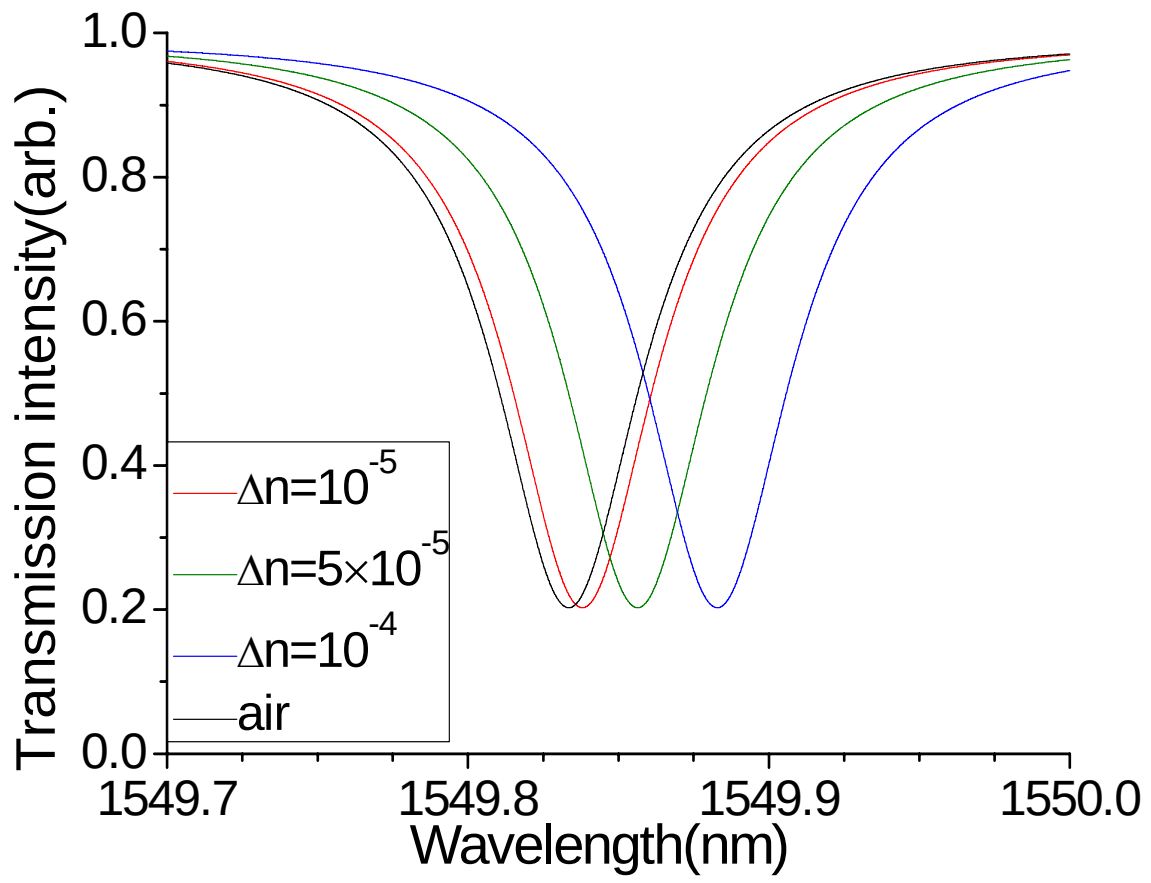

Fig. 5 Sensing an extreme small variation of ambient refractive index.

## 5. Conclusion

We simulate the dependence of sensing performance on the structural parameters of MLRs and conclude that higher $Q$-factor corresponds to large $a$, large $R$ and appropriate $\Delta l$, higher sensitivity which is hardly dependent on $R$ corresponds to small $a$, and lower detection limit corresponds to large $a$, large $R$ and appropriate $\Delta l$. After that we perform the simulation of sensing ambient refractive index according to a set of experimental datas in reference[16] to show the reachable sensing ability of the MLR-based sensor, and the results indicate the detection limit reaches to a variation of $10^{-5}$ RIU. It precedes the demonstrated submicron-diameter optical fiber refractive index sensor. It is indicated that the microfiber loop resonator-based biochemical sensor is a kind of promising photonic device.